\documentclass[a4paper,UKenglish,cleveref, autoref, thm-restate,authorcolumns]{lipics-v2019}

\usepackage{float}
\usepackage[draft]{fixme}

\title{RUPER-LB: Load balancing embarrasingly parallel applications in unpredictable cloud environments}

\titlerunning{} 

\author{V. Gim\'enez-Alventosa}{Instituto de Instrumentac\'ion para Imagen Molecular (I3M), Centro mixto CSIC - Universitat Polit\`ecnica de Val\`encia, Cam\'i de Vera s/n, 46022, Val\`encia, Spain}{vicent.gimenez@i3m.upv.es}{https://orcid.org/0000-0003-1646-6094}{}

\author{Germán Moltó}{Instituto de Instrumentac\'ion para Imagen Molecular (I3M), Centro mixto CSIC - Universitat Polit\`ecnica de Val\`encia, Cam\'i de Vera s/n, 46022, Val\`encia, Spain}{gmolto@dsic.upv.es}{https://orcid.org/0000-0002-8049-253X}{}

\author{J. Dami\'an Segrelles}{Instituto de Instrumentac\'ion para Imagen Molecular (I3M), Centro mixto CSIC - Universitat Polit\`ecnica de Val\`encia, Cam\'i de Vera s/n, 46022, Val\`encia, Spain}{dquilis@dsic.upv.es}{https://orcid.org/0000-0001-5698-7965}{}

\authorrunning{V. G. A. , G. M. M. and J. D. S. Q.} 

\Copyright{Vicent G. A. , Germ\'an M. M. and J. Dami\'an S. Q.} 



\category{} 

\relatedversion{} 

\supplement{}

\acknowledgements{This study was supported by the program ``Ayudas para la contratación de personal investigador en formación de carácter predoctoral, programa VALi+d'' under grant number ACIF/2018/148 from the Conselleria d’Educació of the Generalitat Valenciana and the ``Fondo Social Europeo'' (FSE). The authors would like to thank the Spanish "Ministerio de Econom\'ia, Industria y Competitividad" for the project ``BigCLOE'' with reference number TIN2016-79951-R.}

\nolinenumbers 

\hideLIPIcs  

\begin{document}

\maketitle
\begin{abstract}
The suitability of cloud computing has been studied by several authors to run scientific applications. However, the unpredictable performance fluctuations in these environments hinders the migration of scientific applications to cloud providers. To mitigate these effects, this work presents RUPER-LB, a load balancer for loosely-coupled iterative parallel applications that runs on infrastructures with disparate computing capabilities. The results obtained with a real world simulation software, show the suitability of RUPER-LB to adapt this kind of applications to execution environments with variable performance and highlight the convenience of its adoption.
\end{abstract}

\section{Introduction}
\label{sec:Introduction}

Since the emerging of cloud computing, several authors have studied its suitability  to run scientific applications. The motivation of these studies are the inherent benefits offered by cloud providers. First, cloud computing allows to scale the underlying infrastructure to fit the user needs, eliminating the effects of both under and over provisioning resources. Then, the pay-per-use model provides a cost-effective usage of resources, allowing the users to deploy the required infrastructure and pay for it only during the execution time. Finally, virtualisation provides increased flexibility, since Virtual Machines (VM) can be configured with all the dependencies required by the applications.

However, clouds are not widely used for all kind of scientific applications because they also exhibit some drawbacks. First, cloud providers use a multi-tenant approach to optimise resource usage. This means that the physical processors, disk, memory, etc. where the VM is running can be shared with VMs from another user. This hardware sharing causes a variability on the CPU performance, memory bandwidth, network communications and disk I/O speed, a problem commonly known as \textit{noisy neighbour} \cite{8102951}. In addition, cloud providers typically offer instance types featuring certain characteristics, such as amount of RAM, number of virtual equivalent CPUs (vCPUs), storage, etc., but the user cannot select the specific hardware characteristics. These vCPUs are not physical cores, but a CPU equivalent unit. Unfortunately, the performance of these vCPUs are highly dependent on the underlying hardware, which produce high performance differences between instances of the same type. All these effects have been widely studied in the bibliography \cite{5948601, 10.1145/2885497, 10.14778/1920841.1920902, doi:10.1080/02564602.2017.1393353} and even methodologies are provided to correctly measure this variability \cite{10.1145/3030207.3030229}.

As a response to the demand of instances with predictable capabilities, some providers such as Amazon Web Services (AWS) offer the option to launch single-tenant instances \cite{AWSsingleTenant} at the expense of additional costs. However, depending on the application this fee may not be worth. Also, these single-tenant instances ensure that the physical hardware will be used only by VMs from the account owner. However, this does not preclude from suffering noisy neighbour effects among the user's own instances.

Turning to parallel scientific applications, their execution time is usually determined by the slowest process, so an unbalanced situation will delay the entire application. These facts highlight the need for advanced load balancing techniques to adapt scientific applications to the variable performance found on heterogeneous environments. This effort has been done for High Performance Computing (HPC) applications where authors have studied the suitability of cloud computing environments \cite{5708447} \cite{6200551} \cite{6753812}. These studies agree that tightly coupled applications are less suitable for cloud computing, which is reasonable considering the fluctuations reported on network bandwidth. To mitigate the unbalance problem, several load balancing algorithms adapted to cloud environments have been proposed \cite{6337481} \cite{6546119}. In addition, we can find studies of techniques for efficient VM deployment \cite{7274674} \cite{6253521}. However, this unpredictable variability of the computational capabilities does not only affect tightly coupled processes, but also loosely coupled ones. 

Loosely coupled applications neither require a continuous communication nor synchronisation points, like HPC applications. For instance, most of the load balancing algorithms designed for HPC involve an unnecessary overhead for these applications due the amount of synchronisation points and communications involved. On the other hand, classic load balancing algorithms used on heterogeneous systems, which rely on previous knowledge of the underlying performance \cite{10.1093/comjnl/40.6.356}, are not suitable for these environments due the unpredictable performance fluctuations.

To address these problems, we present RUPER-LB (Runtime Unpredictable Performance Load Balancer) a load balancing algorithm for loosely coupled applications running on environments with unpredictable performance variability with both multi-process and multi-thread balance. RUPER-LB is provided as open-source code under the GPLv3 license and can be download from https://github.com/PenRed/RUPER-LB. For assessment purposes, RUPER-LB was used to balance PenRed \cite{PenRed} simulations, which is a radiation transport simulation framework focused on medical applications with MPI and multithreading built-in parallelism.

\section{Materials and Methods}
\label{sec:MandM}

RUPER-LB focuses on parallel iterative applications such as Monte-Carlo simulations, iterative solvers or multi-parametric analysis. These applications must comply with the following restrictions:

Firstly, the application must be split in tasks. During the execution of these tasks, the application should not require any communication or synchronisation point among the executing threads or processes. Nevertheless, if communications are required, their overhead on the task performance should be negligible. If these assumptions are not accomplished, RUPER-LB can still be used but an HPC-like load balancing algorithm may achieve better results in terms of makespan.

Secondly, the application should measure its speed at runtime. Thus, RUPER-LB assumes that the application behaves like an iterative process, whose speed is measured in iterations per second. The number of iterations to process by each thread and process should be allowed to be changed at runtime. Notice that RUPER-LB neither requires an homogeneous computational cost for the iterations nor a previous balanced distribution among threads.

PenRed, the selected code to test the presented algorithm, satisfies these required assumptions. In this code, tasks correspond to each particle source defined by the user. Each generated primary particle and all its secondaries will be considered as a single history, which corresponds to one iteration. Finally the number of histories to simulate by each thread and process can be changed at runtime.

\subsection{Multi-threading balance}
\label{sec:mthBalance}

Some multi-threading applications employ the involved threads in an unbalanced way. For example, assigning I/O operations or network communications to a specific thread. Also, the computational cost of the iterations that constitute the process could be heterogeneous, or some thread could use accelerated hardware like a GPGPU. Both situations will produce variable unbalances on thread speeds, measured in iterations per second. Also, it is not feasible in a Cloud to know which computational resources are being shared with other VMs running on the same physical hardware and, therefore, how their workload pattern will change during the execution. This fact could increase the unbalance produced by previous effects. Thus, we need to balance the workload between the threads of a single process dynamically. This section describes how this local load balancing is performed.

\begin{figure}
    \centering
    \includegraphics[scale=0.50]{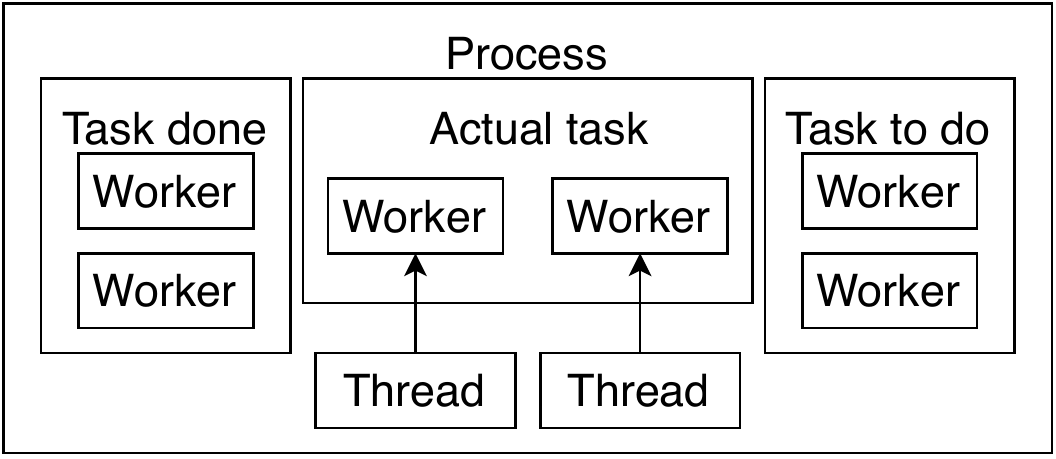}\vspace{0.2cm}
    
    \includegraphics[scale=0.50]{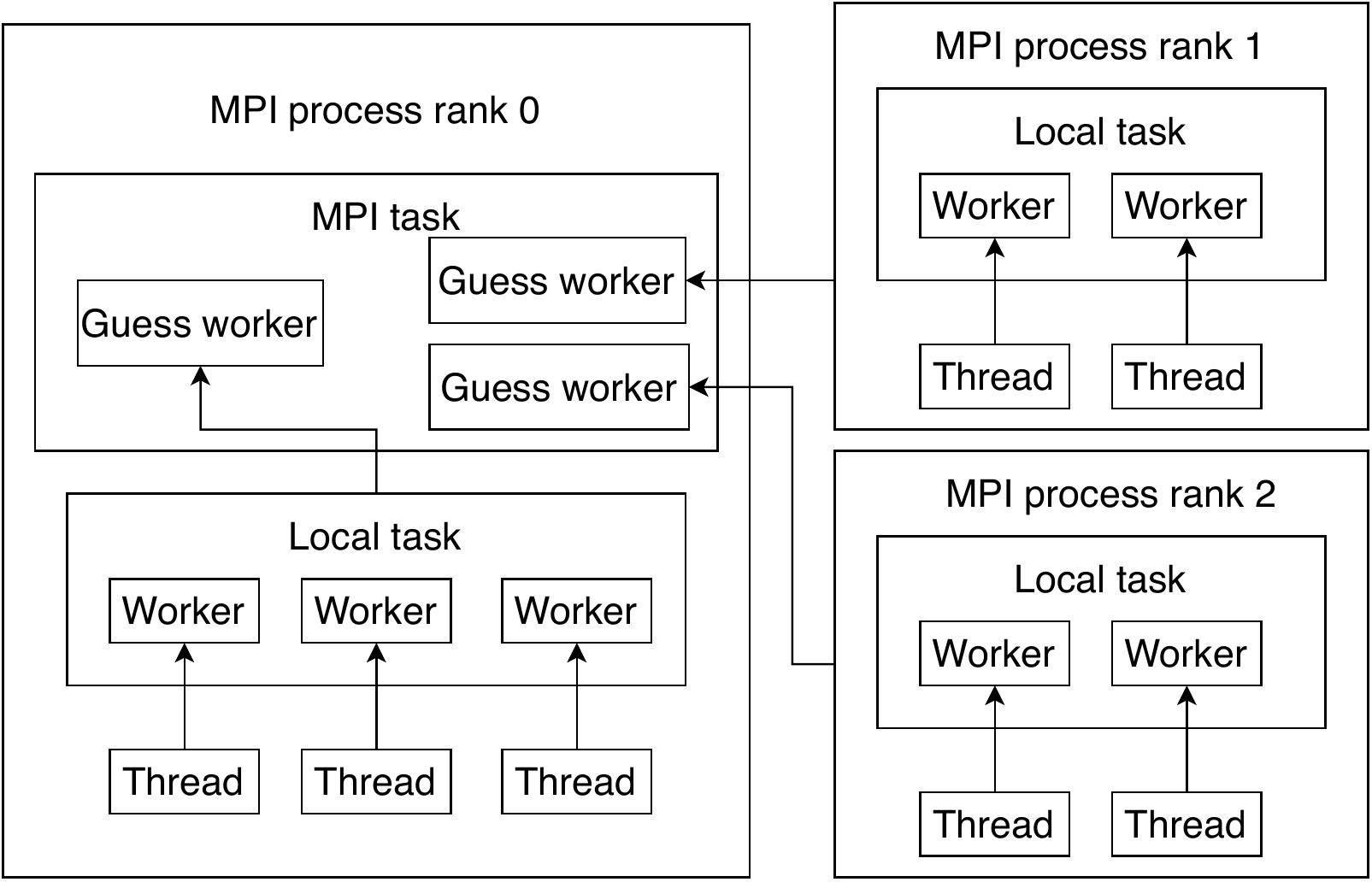}
    \caption{Top: Thread balance system schema for a process with $3$ tasks. Bottom: MPI balancing schema for $3$ MPI processes and $1$ task.}
    \label{fig:BalanceScheme}
\end{figure}

The workload distribution, i.e. the number of iterations assigned to each thread, is handled by two components implemented as classes in an object oriented programming (OOP) language. These are the {\it tasks} and the {\it workers}, which represent a single task and the threads executing the task respectively. Also, the execution could involve more than one task, each of them having its own workers. Figure \ref{fig:BalanceScheme} top shows the basic balance schema for single process executions, where each thread is assigned to a single worker of the active task. The basic states of both components are listed in table \ref{tab:basicStates}.

\begin{table}
    \centering
    \caption{Worker (left) and task (right) object states.}
    \begin{tabular}{|c|p{4cm}|} \hline
        {\bf Variable} & {\bf Description} \\\hline
        $I_n$ &  Assigned iterations\\
        $started$ &  Flags the task start\\
        $finished$ &  Flags the task end\\
        $I_d$ &  Number of finished iterations\\
        $t_r$ &  Last report timestamp\\
        $t_i$ &  Task start timestamp\\
        $m$ & Velocity measures vector\\
        \hline
    \end{tabular}
    \begin{tabular}{|c|p{4cm}|} \hline
        {\bf Variable} & {\bf Description} \\\hline
        $I_n$ & Number of iterations to do\\
        $w$ &  Vector of {\it worker} objects\\
        $t_0$ &  Task start timestamp\\
        $t_{pc}$ & Last checkpoint timestamp\\
        $\Delta t_{pc}$ & Time between checkpoints\\
        $started$ &  Flags task start\\
        $finished$ &  Flags task finish\\
        $t_{min}$ & Balance time threshold\\
        $ds_{max}$ & Maximum speed deviation\\
        \hline
    \end{tabular}
    \label{tab:basicStates}
\end{table}

Basically, each worker reports periodically the number of completed iterations to the {\it task} object. This is done using the {\it report} method, whose code is shown in figure \ref{fig:local-reportAndAddMeasure} left. In this code, and the following ones, the use of locks and the sanity checks on variable values have been omitted for simplicity. The {\it report} method takes as argument three values: a measure of the number of completed iterations, the measure timestamp  and the worker index that performed these iterations. Regarding the execution, first, we use two auxiliary {\it worker}'s methods, {\it working} and {\it elapsed}. The first one returns {\it true} if the {\it worker} is still executing the task, otherwise returns {\it false}, and the second one returns the elapsed time since the last report. Following, the {\it worker} method {\it addMeasure} (Figure \ref{fig:local-reportAndAddMeasure} right) is used to compute and store its speed measured since the last report ($t_r$). In addition, that method returns the quotient $s/s_l$, where $s$ is the new speed to register and $s_l$ is the registered speed in the previous report, that is, the speed deviation from the previous report. This information will be used to calculate, in the {\it report} method, the suggested time interval until next report ($\Delta t$).

\begin{figure}
    \centering
    \begin{tabular}{|l|} \hline
        \multicolumn{1}{|c|}{$report(i, I_{done},t)$} \\\hline
        {\bf Input:}\\
         $i \rightarrow$ Worker index\\
         $I_{done} \rightarrow$ Number of completed iterations\\
         $t \rightarrow $ Report timestamp\\
        {\bf Output:}\\
        $\Delta t \rightarrow$ Suggested time until next report\\\hline
        $ {\bf if}\;  w_i.working()\; {\bf then}$\\
        $\;\;\; \Delta t \leftarrow w_i.elapsed(t)$\\
        $\;\;\; dev \leftarrow w_i.addMeasure(t,I_{done})$\\
        $\;\;\; dev \leftarrow ABS(dev-1)$\\
        $\;\;\;  {\bf if}\;  dev > ds_{max}\; {\bf then}$\\
        $\;\;\;\;\;\; \Delta t \leftarrow \Delta t \cdot max(1-(dev-ds_{max}),0.8)$\\
        $\;\;\;  {\bf else\; if}\;  dev < 0.1 \cdot ds_{max}\; {\bf then}$\\
        $\;\;\; \;\;\; \Delta t \leftarrow \Delta t \cdot min(1+(0.5 \cdot ds_{max}-dev),1.2)$\\
        $\;\;\;  {\bf end\; if}$\\
        $\;\;\;  {\bf if}\;  \Delta t > \Delta t_{pc}\; {\bf then}$\\
        $\;\;\; \;\;\; \Delta t \leftarrow \Delta t_{pc} \cdot 0.8$\\
        $\;\;\;  {\bf end\; if}$\\
        ${\bf else}$\\
        $\;\;\; \Delta t \leftarrow -1$\\
        ${\bf end\; if}$\\
        \hline
    \end{tabular}
    \begin{tabular}{|l|} \hline
        \multicolumn{1}{|c|}{$addMeasure(t,I_{done})$} \\\hline
        {\bf Input:}\\
           $I_{done} \rightarrow$ Number of completed\\
           $\;\;\;\;\;\;\;\;\;\;\;$ iterations\\
           $t \rightarrow$ Measure timestamp\\
        {\bf Output:}\\
          $dev \rightarrow $ Speed deviation\\
          \hline
        $\Delta t \leftarrow t-t_r$\\
        $\Delta t_m \leftarrow t-t_i$\\
        $\Delta I \leftarrow I_{done}-I_d$\\
        $s_l \leftarrow speed()$\\
        $s \leftarrow \Delta I / \Delta t$\\
        $I_d \leftarrow I_{done}$\\
        $t_r \leftarrow t$\\
        $dev \leftarrow s/s_l$\\
        $m \leftarrow (\Delta t_m, s)$\\
        \hline
    \end{tabular}    
    
\caption{{\it Task} {\it report} method (left) and {\it Worker} {\it addMeasure} method (right).} 
\label{fig:local-reportAndAddMeasure} 
\end{figure} 

Each thread will compute its own reports independently, i.e. the threads do not require to synchronise to perform the report at the same time. The same goes for the {\it checkpoint} method, whose pseudocode is shown in figure \ref{fig:checkpoint-addMeasureGuess} left. This {\it task} method, redistributes the workload among its {\it workers} according to the information stored by reports. First of all, the algorithm calculates three values: the total simulation speed ($s_t$), the total reported iterations done ($I_t$) and the predicted iterations done ($I_{pred}$). To obtain $I_{pred}$, we use the auxiliary {\it worker} method {\it predDone}, which returns the predicted iterations done by the worker assuming no changes on its speed since last report. Notice that the calculation of task speed excludes the already finished workers. Then, we check if the required iterations have been done. If that happens, the assigned iterations of each worker will be set to its reported iterations done, i.e. force workers to finish the task. On the other hand, if there are still iterations to do, we evaluate a prediction of the remaining execution time ($t_{res}$) according to $I_{pred}$ and $s_t$. Finally, if $t_{res}$ is greater than the threshold ($t_{min}$), the iterations assigned to each active worker will be recalculated according to its speed factor.

At some point of the execution, the workers will consider that they have finished the task. At this point, workers will ask to finish to the {\it task} object, which will allow or refuse the request to finish according to the {\it task} stored information. There are two reasons to deny this request. The first reason is that the {\it task} object has registered less iterations done by the worker than the ones assigned. In this case, a new report will be required. The second reason is that the estimated remaining execution time to complete the task is greater than $t_{min}$. This last case requires a new checkpoint to reassign the number of iterations for each worker. If neither of both conditions are accomplished, the worker can finish the task. Thus, the {\it worker} method {\it working} will return {\it false} hereinafter. Once all workers have finished, the task is considered as finished.

\begin{figure}
    \centering
    \begin{tabular}{|l|} \hline
        \multicolumn{1}{|c|}{$checkPoint()$} \\\hline
        {\bf Input:}\\
        {\bf Output:}\\
        \hline
        $t_{pc} \leftarrow actualTime()$\\
        $s_{t} \leftarrow 0$\\
        $I_{t} \leftarrow 0$\\
        $I_{pred} \leftarrow 0$\\
        ${\bf for\; each}\; worker \; {\bf in}\; w \; {\bf do}$\\
        $\;\;\; I_{t} \leftarrow I_{t} + worker.I_d$\\
        $\;\;\; {\bf if}\; worker.working() \; {\bf then}$\\
        $\;\;\;\;\;\; s_{t} \leftarrow s_{t} + worker.speed()$\\
        $\;\;\;\;\;\; I_{pred} \leftarrow I_{pred} + worker.predDone(t)$\\
        $\;\;\; {\bf else}$\\
        $\;\;\;\;\;\; I_{pred} \leftarrow I_{pred} + worker.I_d$\\
        $\;\;\; {\bf end\; if}$\\
        $ {\bf end\; for}$\\
        $ {\bf if}\; I_n <= I_t\; {\bf then}$\\
        $\;\;\; {\bf for\; each}\; worker \; {\bf in}\; w \; {\bf do}$\\
        $\;\;\;\;\;\; {\bf if}\; worker.working() \; {\bf then}$\\
        $\;\;\;\;\;\;\;\;\; worker.I_n \leftarrow worker.I_d$\\
        $\;\;\;\;\;\; {\bf end\; if}$\\
        $\;\;\; {\bf end\; for}$\\
        $ {\bf else}$\\
        $\;\;\; I_{res} \leftarrow I_n - I_{pred} $\\
        $\;\;\; t_{res} \leftarrow I_{res} / s_{t} $\\
        $\;\;\; {\bf if}\; t_{res} > t_{min}\; {\bf then}$\\
        $\;\;\;\;\;\; {\bf for\; each}\; worker \; {\bf in}\; w \; {\bf do}$\\
        $\;\;\;\;\;\;\;\;\; {\bf if}\; worker.working() \; {\bf then}$\\
        $\;\;\;\;\;\;\;\;\;\;\;\; s_{fact} \leftarrow  worker.speed()/s_t $\\
        $\;\;\;\;\;\;\;\;\;\;\;\; worker.I_n \leftarrow worker.I_d \;+$\\
        $\;\;\;\;\;\;\;\;\;\;\;\;\;\;\;\;\;\;\;\;\;\;\;\;\;\;\; s_{fact} \cdot (I_n-I_t) $\\
        $\;\;\;\;\;\;\;\;\; {\bf end\; if}$\\
        $\;\;\;\;\;\; {\bf end\; for}$\\
        $\;\;\; {\bf end\; if}$\\
        $ {\bf end\; if}$\\
        \hline
    \end{tabular}
	\begin{tabular}{|l|} \hline
		\multicolumn{1}{|c|}{$addMeasure(t, I_{done})$} \\\hline
		{\bf Input:}\\
		$I_{done} \rightarrow$ Iterations completed\\
		$\;\;\;\;\;\;\;\;\;\;\;$ prediction\\
		$t \rightarrow$ Measure timestamp\\
		{\bf Output:}\\
		$dev \rightarrow $ Speed deviation\\
		\hline
		${\bf if}\; speed() = 0\; {\bf then}$\\
		$\;\;\; dev \leftarrow worker::addMeasure(t,I_n)$\\
		${\bf else}$\\
		$\;\;\; \Delta t \leftarrow t-t_r$\\
		$\;\;\; \Delta t_m \leftarrow t-t_i$\\
		$\;\;\; {\bf if}\; I_d > I_{done}\; {\bf then}$\\
		$\;\;\;\;\;\; \bar{s_1} \leftarrow I_d/(t_r-t_i)$\\
		$\;\;\;\;\;\; \bar{s_2} \leftarrow I_{done}/(t-t_i)$\\
		$\;\;\;\;\;\; dev \leftarrow \bar{s_2}/\bar{s_1}$\\
		$\;\;\; {\bf else}$\\
		$\;\;\;\;\;\; \Delta {I_e} \leftarrow speed() \cdot \Delta t$\\
		$\;\;\;\;\;\; \Delta {I_r} \leftarrow I_{done} - I_d$\\
		$\;\;\;\;\;\; dev \leftarrow \Delta {I_r}/\Delta {I_e}$\\
		$\;\;\; {\bf end\; if}$\\
		$\;\;\; s \leftarrow dev \cdot speed()$\\
		$\;\;\; t_r \leftarrow t$\\
		$\;\;\; m \leftarrow (\Delta t_m,s)$\\
		${\bf end\; if}$\\
		\hline
	\end{tabular}
\caption{Method {\it checkPoint} for {\it task} object (left) and {\it addMeasure} for {\it guess worker} object (right).} 
\label{fig:checkpoint-addMeasureGuess} 
\end{figure} 

\subsection{MPI balance}

If MPI load balancing is enabled, this is handled at two levels, as shown in figure \ref{fig:BalanceScheme} bottom. First, locally to each MPI process, where the threads are balanced using the method described in the previous section. Then, the number of iterations to do is split between MPI processes. The rank $0$ will handle the assignment of iterations for each process {\it task}, thus the $I_n$ value is not constant on MPI. For that purpose, both objects {\it worker} and {\it task} are extended as follows. First, since the local thread reports are performed asynchronous, the iterations done and speed registered at local tasks are, in general, outdated. To counteract that, the MPI balance procedure registers the predicted iterations done, and not the reported ones. This procedure requires a new type of worker, which has been created as a derived object of the {\it worker} saw at section \ref{sec:mthBalance}. That new worker object used for MPI balance has been named {\it guess worker}, which shares the same state as the base {\it worker} class (table \ref{tab:basicStates}). However, notice that {\it guess workers} do not represent a single thread, as the workers of section \ref{sec:mthBalance}. Instead, a {\it guess worker} registers the information of the whole task running on one of the MPI processes (figure \ref{fig:BalanceScheme}). In addition, a {\it guess worker} object uses a different {\it addMeasure} method, whose pseudocode is shown in figure \ref{fig:checkpoint-addMeasureGuess} (right). This {\it addMeasure} method corrects the last measured speed using the deviation between the reported and the expected prediction of iterations done at the time $t$. Notice that this method based on speed correction could fail if $0$ iterations per second is reported. To handle this situation, the {\it addMeasure} method of the base {\it worker} object (figure \ref{fig:local-reportAndAddMeasure}) will be called. 

On the other hand, to adapt {\it task} objects to handle MPI balance, we add the variables listed in table \ref{tab:MPItaskState} to its state. As indicated in the following descriptions, the usage of the new variables depends on the MPI process rank. For example, as shown in figure \ref{fig:BalanceScheme}, only the rank $0$ uses the vector $w^{MPI}$ to save the local task reports.

\begin{table}[H]
	\centering
	\caption{MPI {\it task} state extension.} 
	\begin{tabular}{|c|p{8cm}|} \hline
		{\bf Variable} & {\bf Description} \\\hline
		$w^{MPI}$ & Vector of {\it guess workers}. Stores one for each MPI process.\\
		$finished^{MPI}$ & Flags MPI balancing finish\\ 
		$I_n^{MPI}$ & Iterations to do between all MPI processes\\
		$finish_{req}^{MPI}$ & Flags MPI finish request\\
		$finish_{sent}^{MPI}$ & Flags MPI finish request sent\\
		\hline
	\end{tabular}
	\label{tab:MPItaskState}
\end{table}

With these modifications, the report and balance steps are handled by a single thread in each MPI process via the {\it monitor} method. This one has a different behaviour regarding its rank number, as shown in figure \ref{fig:monitor}. Both are explained below.

For rank $0$ (figure \ref{fig:monitor} left), $\Delta t_i^{report}$ and $\Delta t_i^{next}$ save, respectively, the elapsed time between reports and the time until next report for the {\it guess worker} number $i$. Then, {\it receiveAny} waits until some request is received, regardless the origin rank, or until the elapsed time reaches the {\it timeout}. In both cases, the elapsed time will be stored at $\Delta t$. If a request is received, it is stored at {\it req}. After the {\it receiveAny} call, the time until the next report request for each MPI process will be updated according to $\Delta t$. Also, if $\Delta t >= \Delta t_i^{next}$, a report will be requested to the process with rank $i$. Already sent report requests are flagged with $\Delta t_i^{next} = 0$. Finally, the timeout is set to the minimum value in the $\Delta t^{next}$ array.

Regarding the procedure to handle the requests, there exists three possible requests. The first one, with identifier $0$, handles the workers start petitions. As response to this request, the rank $0$ sends a preliminary iteration assignation that will be updated when the first report is received. This part of the code uses the auxiliary method $done^{MPI}()$, which returns the number of the predicted iterations done by all the MPI processes.

The second instruction, with identifier $1$, handles the reception of the reports. For that purpose, the method {\it receiveReport} is used to handle the petition. The functionality of {\it receiveReport} is very similar to the already shown methods {\it report} and {\it checkpoint}, except that it works with predictions of the computed iterations via the {\it guess worker} {\it addMeasure} method. So, it stores the new measure, updates the iteration assignment for MPI workers, and sends to the rank $i$ its new assignation together with a flag to indicate if the MPI balance continues or finishes. As local balance (section \ref{sec:mthBalance}), this will finish when the predicted remaining time is below the threshold. When the MPI balance finishes, the number of assigned iterations for each MPI process will remain unaltered hereinafter. To save space, the pseudocode of this function is not included at this document. However, the details can be found at the provided source code repository. Finally, once the response has been sent, the corresponding time until the next report and the timeout are updated. 

The last instruction, with identifier $2$, handles the finish requests. Like the method used at section \ref{sec:mthBalance}, MPI workers can request to finish the task, attaching a report to their request. The reasons to send a finish request will be explained at the {\it monitor} description for non zero ranks. For instance, these requests are handled by {\it receiveReport} too. Finally, we check if all workers have been notified that the MPI balance has finished. In this case, the monitor execution ends.

\begin{figure}
	\centering
	\begin{tabular}{|l|} \hline
		\multicolumn{1}{|c|}{$monitor()$} \\\hline
		{\bf Input:}\\
		{\bf Output:}\\
		\hline
		${\bf for}\; i = 0\; {\bf until}\; w^{MPI}.size()-1\; {\bf do}$\\
		$\;\;\; \Delta t^{report}_i \leftarrow \Delta t_{pc}$\\
		$\;\;\; \Delta t^{next}_i \leftarrow 0$\\        
		${\bf end\; for}$\\
		$timeout \leftarrow \Delta t_{pc}$\\
		${\bf while\;} true\; {\bf do}$\\
		$\;\;\; req \leftarrow receiveAny(timeout,\Delta t)$\\
		$\;\;\;timeout \leftarrow 10^9$\\
		$\;\;\; {\bf for}\; i = 0\; {\bf until}\; w^{MPI}.size()-1\; {\bf do}$\\
		$\;\;\;\;\;\; {\bf if}\; \Delta t^{next}_i > 0\; {\bf then}$\\
		$\;\;\;\;\;\;\;\;\; {\bf if}\; \Delta t^{next}_i <= \Delta t\; {\bf then}$\\
		$\;\;\;\;\;\;\;\;\;\;\;\; requireReport(i)$\\
		$\;\;\;\;\;\;\;\;\;\;\;\; \Delta t^{next}_i \leftarrow 0$\\
		$\;\;\;\;\;\;\;\;\;{\bf else}$\\
		$\;\;\;\;\;\;\;\;\;\;\;\; \Delta t^{next}_i \leftarrow \Delta t^{next}_i - \Delta t$\\
		$\;\;\;\;\;\;\;\;\;\;\;\; {\bf if}\; timeout > \Delta t^{next}_i\; {\bf then}$\\
		$\;\;\;\;\;\;\;\;\;\;\;\;\;\;\; timeout \leftarrow  \Delta t^{next}_i$\\
		$\;\;\;\;\;\;\;\;\;\;\;\; {\bf end\; if}$\\
		$\;\;\;\;\;\;\;\;\; {\bf end\; if}$\\
		$\;\;\;\;\;\;{\bf end\; if}$\\
		$\;\;\;{\bf end\; for}$\\
		$\;\;\; {\bf if\;} req \; {\bf then}$\\
		$\;\;\;\;\;\; {\bf if\;} req.instruction = 0\; {\bf then}$\\
		$\;\;\;\;\;\;\;\;\; I_{rem} = I^{MPI}_n - done^{MPI}()$\\
		$\;\;\;\;\;\;\;\;\; req.send(I_{rem}/w^{MPI}.size())$\\
		$\;\;\;\;\;\;\;\;\; \Delta t^{next}_{req.node} \leftarrow \Delta t^{report}_{req.node}$\\
		$\;\;\;\;\;\; {\bf else\; if\;} req.instruction = 1\; {\bf then}$\\
		$\;\;\;\;\;\;\;\;\; \Delta t^{report}_{req.node} \leftarrow receiveReport(req)$\\
		$\;\;\;\;\;\;\;\;\; \Delta t^{next}_{req.node} \leftarrow \Delta t^{report}_{req.node}$\\ 
		$\;\;\;\;\;\;\;\;\; {\bf if\;} timeout > \Delta t^{next}_{req.node} \; {\bf then}$\\
		$\;\;\;\;\;\;\;\;\;\;\;\; timeout \leftarrow  \Delta t^{next}_{req.node}$\\
		$\;\;\;\;\;\;\;\;\; {\bf end \; if}$\\
		$\;\;\;\;\;\; {\bf else\; if\;} req.instruction = 2\; {\bf then}$\\
		$\;\;\;\;\;\;\;\;\; receiveReport(req)$\\
		$\;\;\;\;\;\; {\bf end\; if}$\\
		$\;\;\;\;\;\; {\bf if\;} allFinished()\; {\bf then}$\\
		$\;\;\;\;\;\;\;\;\; {\bf exit}$\\
		$\;\;\;\;\;\; {\bf end\; if}$\\        
		$\;\;\; {\bf end\; if}$\\
		${\bf end\; while}$\\
		\hline
	\end{tabular}
	\begin{tabular}{|l|} \hline
		\multicolumn{1}{|c|}{$monitor()$} \\\hline
		{\bf Input:}\\
		{\bf Output:}\\
		\hline
		$I_n \leftarrow send(0)$\\
		${\bf while}\; true\; do$\\
		$\;\;\; req \leftarrow waitAny(finish^{MPI}_{req})$\\
		$\;\;\; {\bf if}\; req\; {\bf then}$\\
		$\;\;\;\;\;\; {\bf if}\; req.instruction = 1 \; {\bf or}\; 2\; {\bf then}$\\
		$\;\;\;\;\;\;\;\;\; t \leftarrow actualTime()$\\
		$\;\;\;\;\;\;\;\;\; I^{pred}_d \leftarrow predDone(t)$\\
		$\;\;\;\;\;\;\;\;\; req.send(t,I^{pred}_d)$\\
		$\;\;\;\;\;\;\;\;\; (I_n, finished^{MPI}) \leftarrow req.receive()$\\
		$\;\;\;\;\;\;\;\;\; {\bf if\;} finished^{MPI}\; {\bf then}$\\
		$\;\;\;\;\;\;\;\;\;\;\;\; {\bf exit}$\\
		$\;\;\;\;\;\;\;\;\; {\bf end\; if}$\\
		$\;\;\;\;\;\;\;\;\; {\bf if}\; req.instruction = 2\; {\bf then}$\\
		$\;\;\;\;\;\;\;\;\;\;\;\; finish^{MPI}_{sent} \leftarrow false$\\
		$\;\;\;\;\;\;\;\;\; {\bf end\; if}$\\
		$\;\;\;\;\;\; {\bf end\; if}$\\
		$\;\;\; {\bf else}$\\
		$\;\;\;\;\;\; send(2)$\\
		$\;\;\;\;\;\; finish^{MPI}_{req} \leftarrow false$\\
		$\;\;\;\;\;\; finish^{MPI}_{sent} \leftarrow true$\\        
		$\;\;\; {\bf end\; if}$\\
		${\bf end\; while}$\\
		\hline
	\end{tabular}
	\caption{Methods {\it monitor} of the object {\it task} for MPI rank $0$ (left) and greater than zero (right).} 
	\label{fig:monitor} 
\end{figure} 

For the other ranks, which constitute the MPI workers, the monitor pseudocode is shown in figure \ref{fig:monitor} right. First of all, the monitor sends a start petition to the rank $0$ and receives the initial assignation of iterations to do. Once inside the loop, the function {\it waitAny} waits to receive a petition or a response from the rank $0$ or until the value of the variable $finish_{req}^{MPI}$ changes to {\it true}. 

On the first case, whether the received instruction identifier is $1$ or $2$, the monitor sends the predicted computed iterations ($I_d^{pred}$) at time instant $t$. Then, it waits to receive the response of the rank $0$ with the new iteration assignation and the flag to finish the MPI balance ($finished^{MPI}$). If the MPI balancing has finished, the monitor process ends. Finally, if this request is a response of a finish petition (instruction $2$), the $finish_{sent}^{MPI}$ is set to $false$ to allow triggering new finish petitions.

Instead, if $finish_{req}^{MPI}$ has changed its value to $true$, the monitor sends an instruction petition $2$ to ask to finish the MPI balance. Also, the values of the flags $finish_{req}^{MPI}$ and $finish_{sent}^{MPI}$ are changed to $false$ and $true$, respectively. The value of $finish_{req}^{MPI}$ can be changed to $true$ by local threads when they try to finish the task. This happens when a worker satisfies the criteria to finish the local task shown in section \ref{sec:mthBalance}. However, if the MPI balance is still active, the number of iterations to carry out could change. For instance, the local task cannot allow its workers to exit the task. Instead, the local task sends a finish petition to rank $0$. In addition, the flag value could also change when a local {\it checkpoint} call reaches a remaining time lower than the threshold.

\section{Results}
\label{sec:results}

To test the efficiency of the proposed algorithm, we have simulated the variable overhead caused by neighbour VMs on an on-premises cloud managed by OpenStack. Its underlying infrastructure is composed by nodes with two Skylake Gold 6130 at 2.1 GHz with 16 cores each and 768 GB RAM DDR4@2666. 

\begin{figure}
	\centering
	\includegraphics[scale = 0.55]{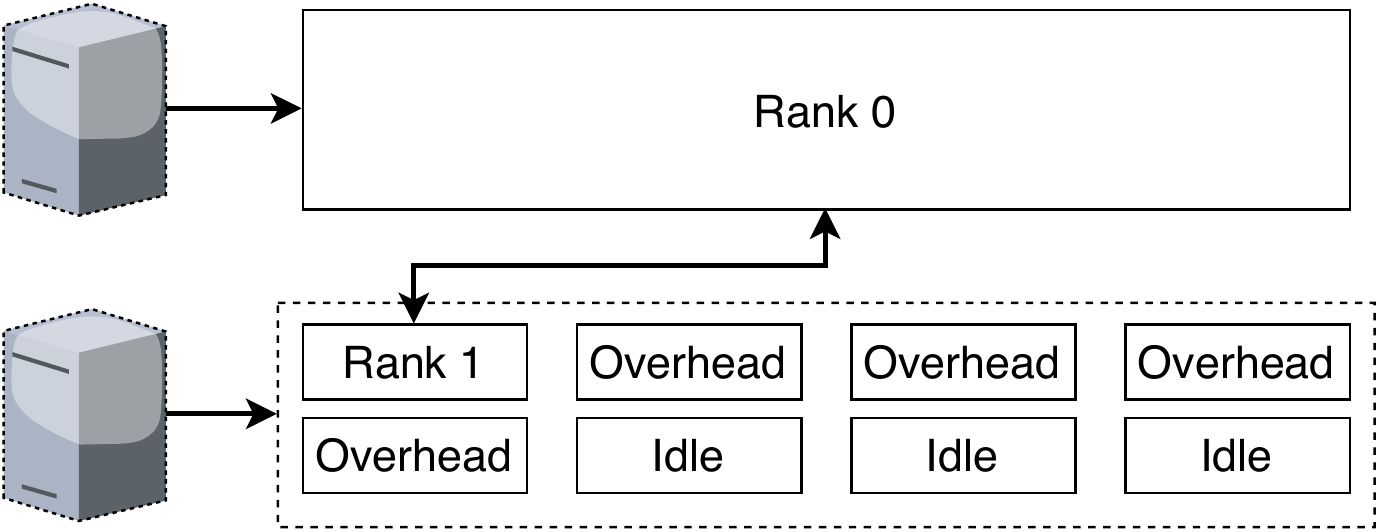}
	\caption{Test infrastructure schema.}
	\label{fig:arqExp}
\end{figure}

The deployed infrastructure for our experimentation consists of two physical nodes, as shown in figure \ref{fig:arqExp}. On the first, a single VM was deployed with $64$ vCPUs to ensure that the physical node is not shared with any other VM. The second one is filled with smaller VMs with $8$ vCPUs each one. On the second node, only one of the small VMs will execute the PenRed simulations. Also, four of the other small VMs, will execute a dummy process whose CPU usage depends on the time of day. These overhead tasks are bash scripts which run the command {\it yes} followed by a {\it sleep}. The sleep time depends, as we said, on the time of day. With this approach, we simulate a variation of the CPU usage of the neighbours VMs. The other VMs remain idle, and their only purpose is to fill the physical node.

Regarding the application to balance, we have selected PenRed \cite{PenRed} code system, which implements the PENELOPE \cite{PENELOPE} physics in an extensible parallel engine for radiation transport in matter simulations. Some of its usages are performing simulations of clinical radiation treatments, radiological protection, or industrial applications. To test RUPER-LB we will use the PenRed simulation example {\it 2-plane}, provided as part of the software distribution.

With that experimental setup, we have executed the very same simulation with and without load balancing. We have configured the minimum time between checkpoints ($\Delta t_{pc}$) to $300\,s$, which has been selected according to process execution time order. Thus, we expect to see executing times delay between ranks and threads lower than $300\,s$. In the following experiments, two MPI processes have been used. The process with rank $0$ runs on the large VM, i.e. with no neighbour influence. Thus, the process with rank $1$ is executed at the node with multiple tenants. In addition, both processes use $8$ threads each.

\begin{figure}
	\centering
	\includegraphics[scale=0.50]{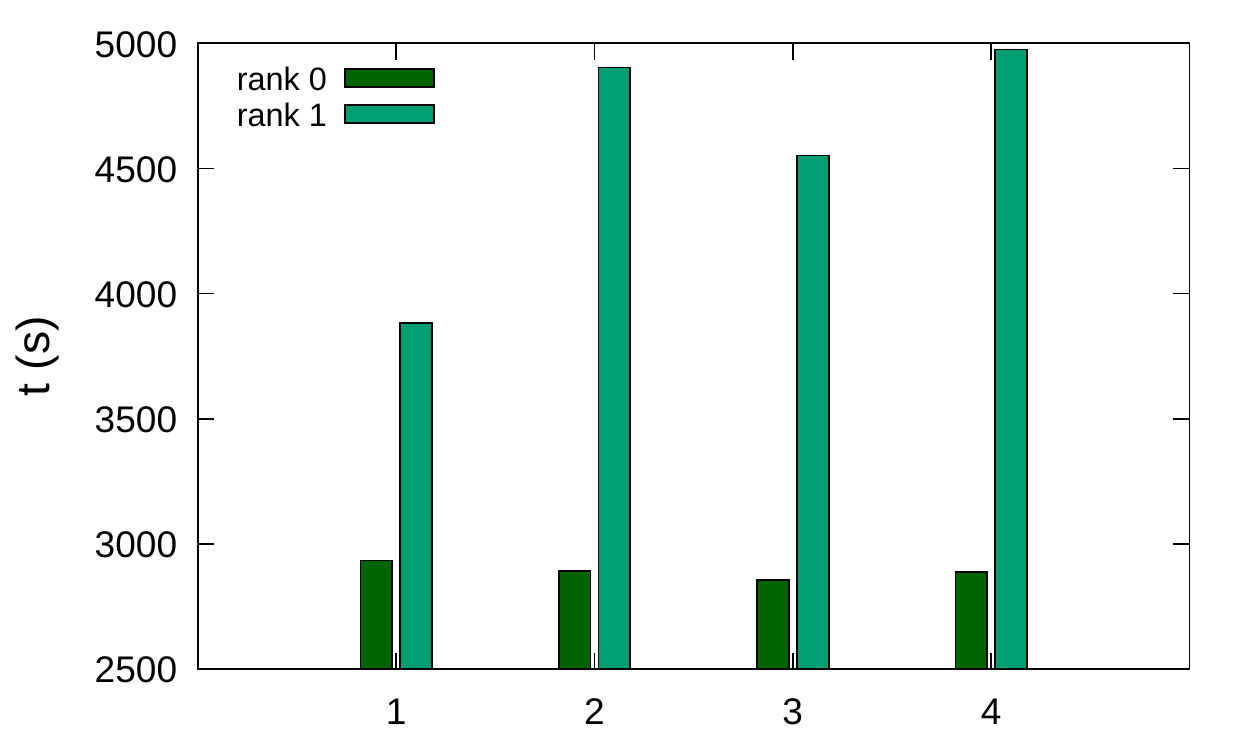}
	\includegraphics[scale=0.50]{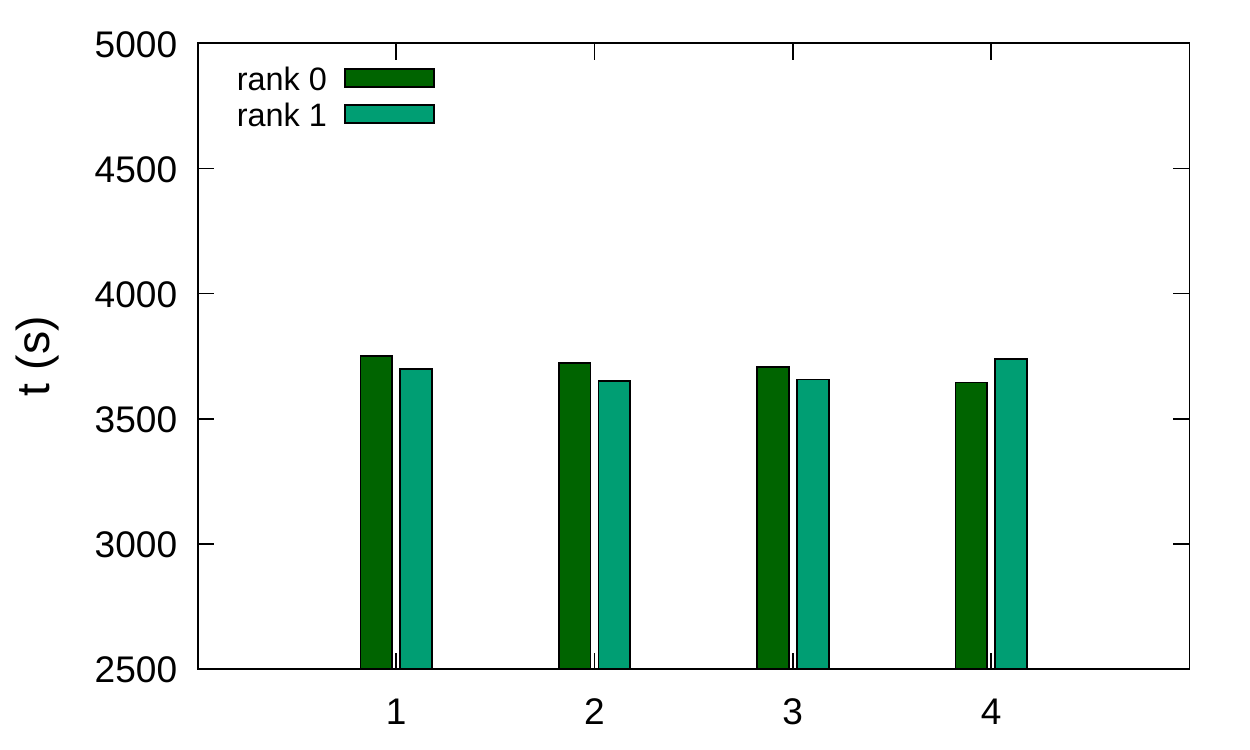}
	\caption{Execution time using $2$ MPI processes with $8$ threads each one. Left: without load balance. Right: with load balance.}
	\label{fig:shortOverload}
\end{figure}

\begin{figure}
    \centering
    \includegraphics[scale=0.50]{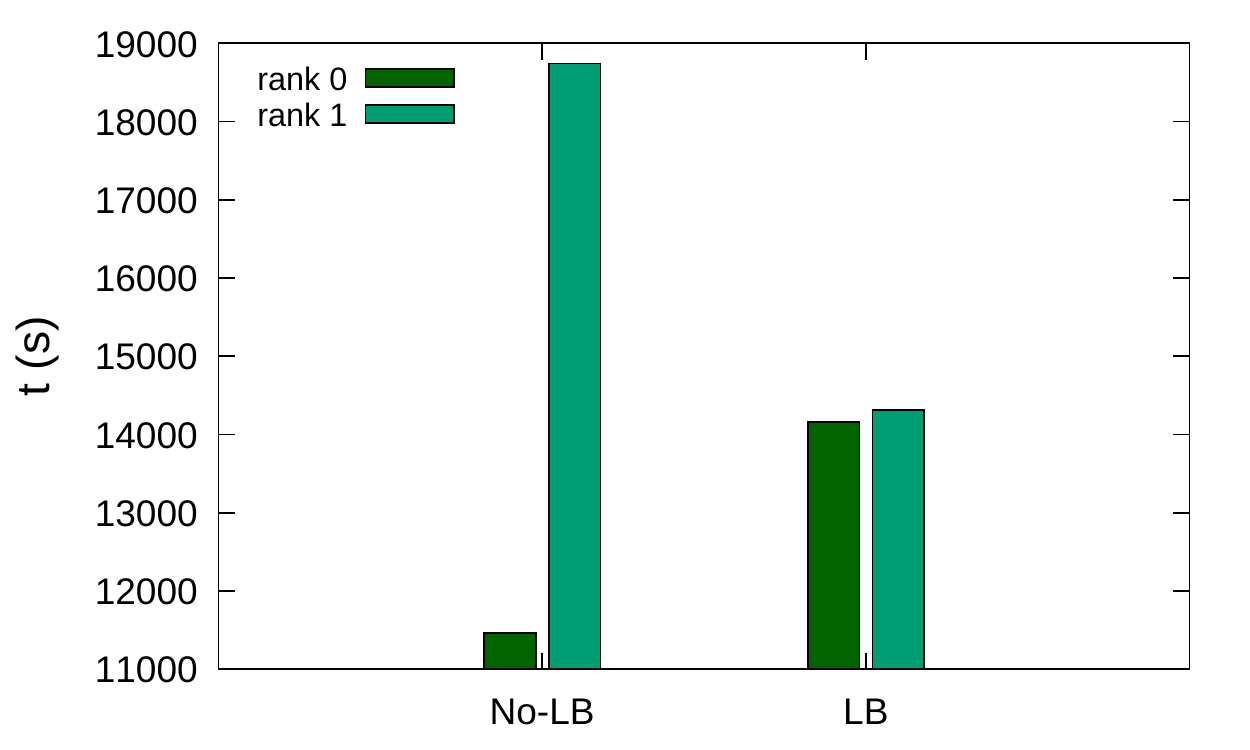}
    \includegraphics[scale=0.50]{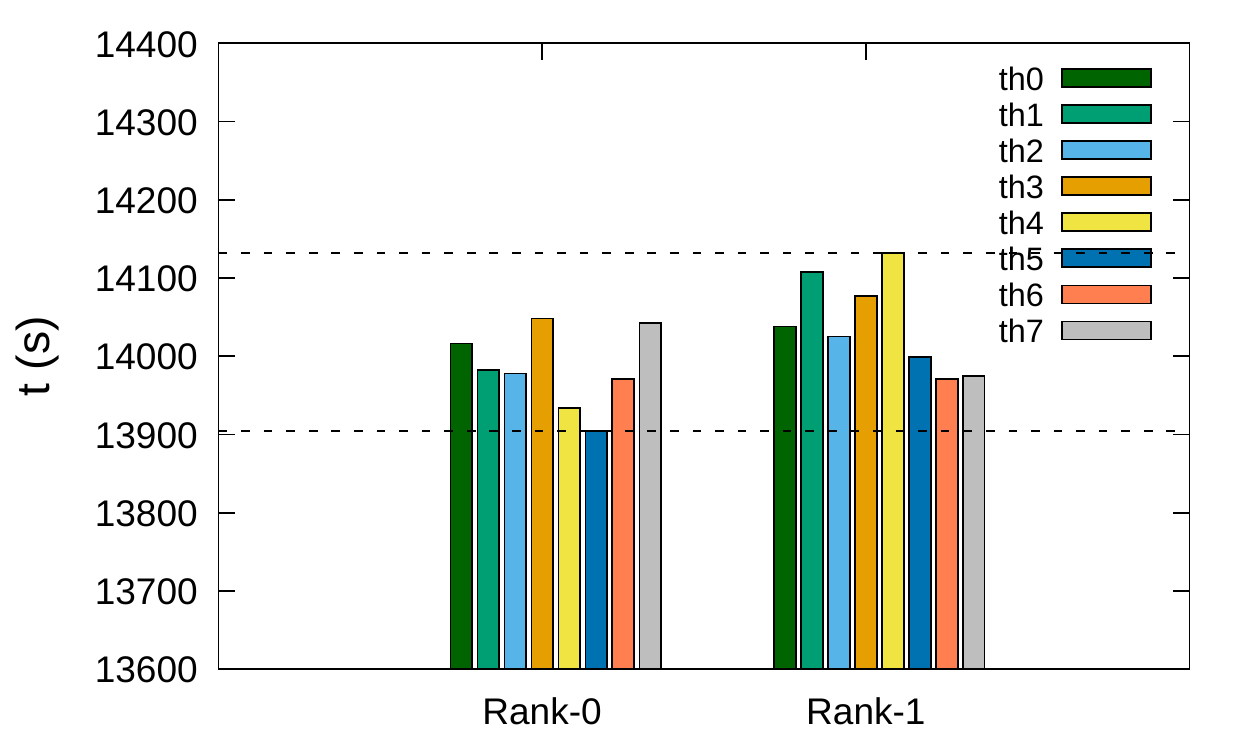}
    \caption{Execution times for simulations with a higher number of iterations, by rank (left) and by thread with load balance (right).}
    \label{fig:longOverload}
\end{figure}

The same simulation was repeated $4$ times both with and without load balancing. Figure \ref{fig:shortOverload} shows the execution time of every process by rank number, for each simulation run. As we can see, on the load balanced results, the delay between ranks is smaller than the selected $\Delta t_{pc}$. At the following test, we have increased the computational cost increasing the number of iterations (Figure \ref{fig:longOverload}). As expected, maintaining the same value of $\Delta t_{pc}$, the relative differences on execution time are reduced. 

For the same simulation, with load balancing enabled, figure \ref{fig:longOverload} right shows the execution time for each thread of each MPI process. There, the dashed lines limits the fastest and the slowest thread for both ranks, and we can check that the corresponding delay is below $\Delta t_{pc}$.

\begin{figure}
    \centering
    \includegraphics[scale=0.50]{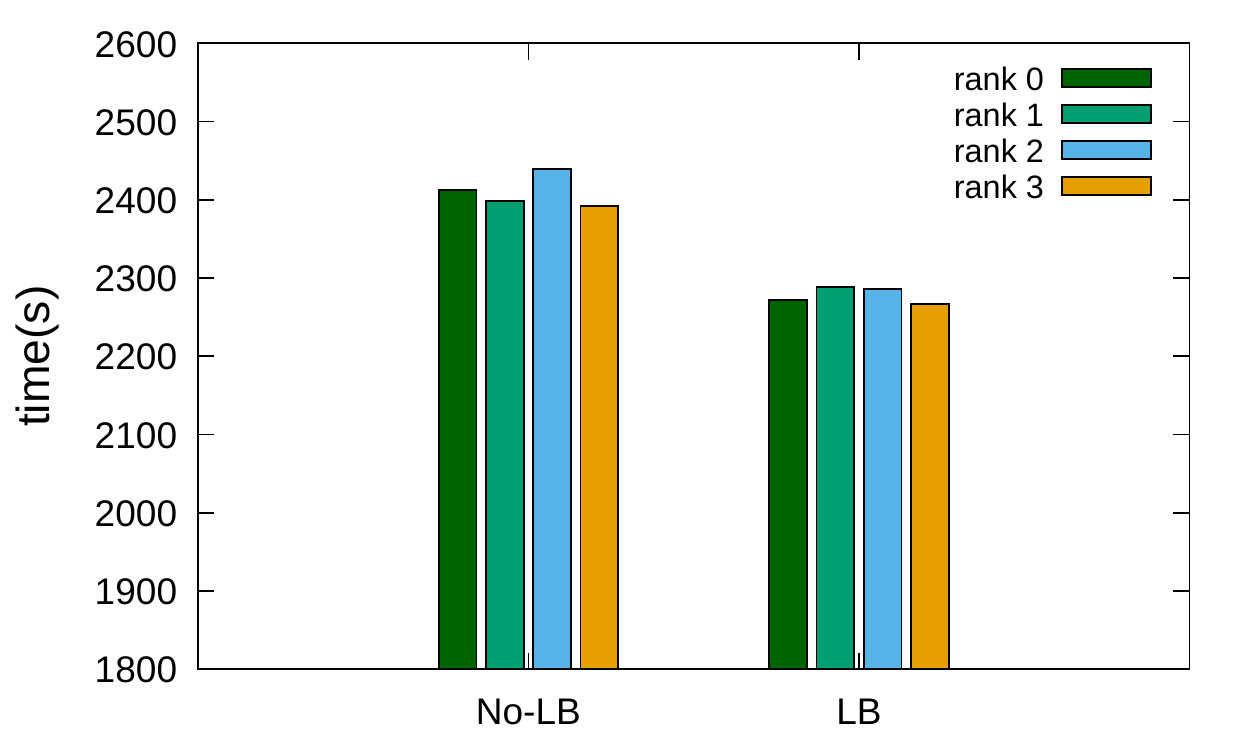}
    \caption{Simulations executed with $4$ MPI processes and $8$ threads each one on the single-tenant node.}
    \label{fig:localNoOverload}
\end{figure}

\begin{figure}
    \centering
    \includegraphics[scale=0.60]{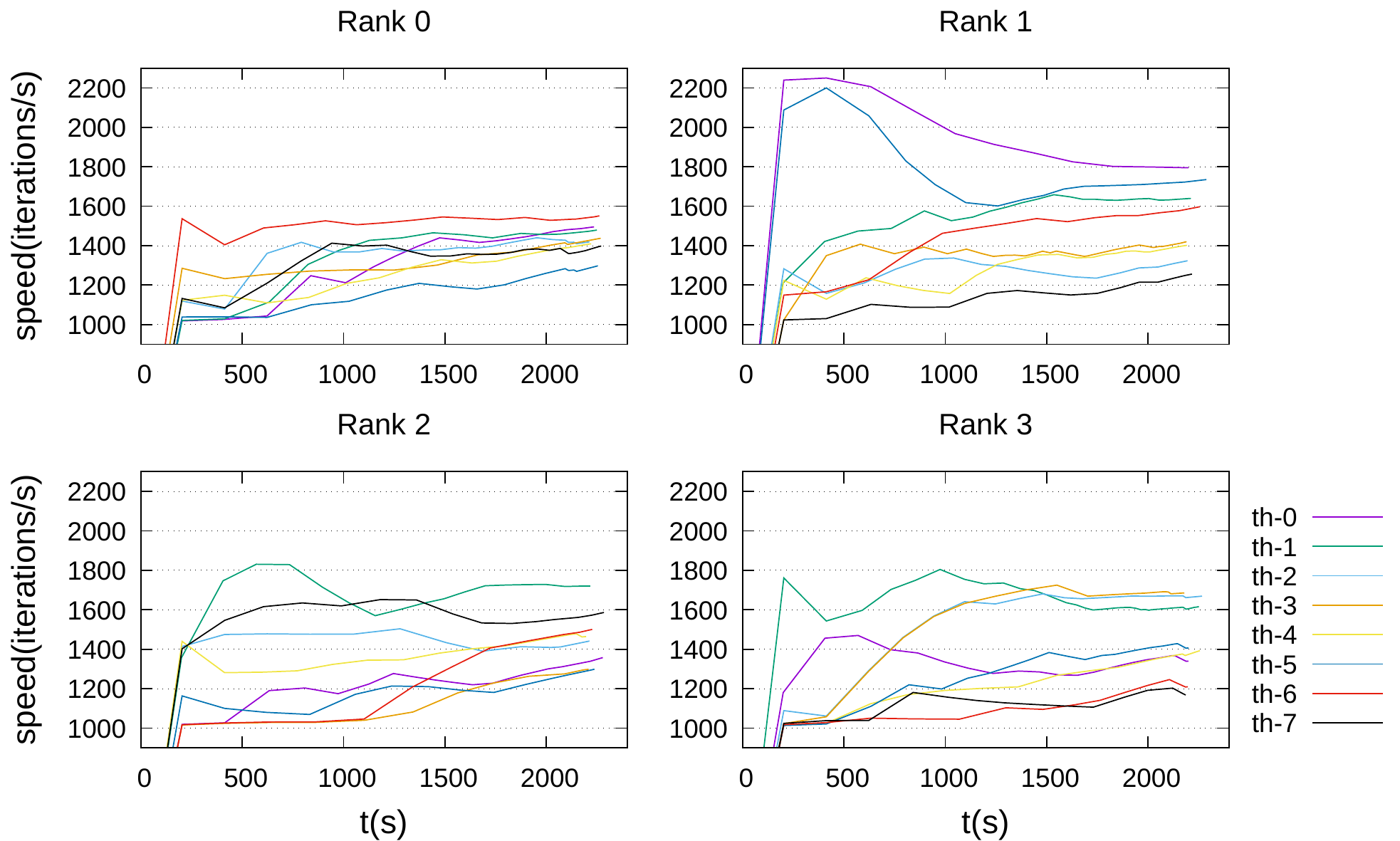}
    \caption{Evolution of the mean speed for each thread in each MPI process.}
    \label{fig:threadUnbalance}
\end{figure}

To test how RUPER-LB can save execution time inside a single node, we have executed the same simulation using $4$ MPI processes with $8$ threads for each one, but all of them running on the single-tenant node. This simulation has been executed with and without load balancing. The corresponding execution times for each rank are shown in figure \ref{fig:localNoOverload}. The same simulation with load balancing enabled is about a $6-7\%$ faster. To understand the results shown in figure \ref{fig:localNoOverload}, we have represented the mean speed evolution of the threads of each MPI process in figure \ref{fig:threadUnbalance}. As we can see, at the end of the execution the mean speeds present non negligible differences between the threads of the same MPI process. This fact explains why RUPER-LB achieves shorter execution times on this test. On the other hand, to explain why figure \ref{fig:localNoOverload} seems to show no unbalance between ranks, notice that the execution time of each rank is determined by the slowest thread. Even if there exists unbalance between the threads, if the slowest thread of each rank requires approximately the same execution time in all of them, that gives the false appearance that the whole process is well balanced. 

\section{Conclusions}
\label{sec:conclusions}

This work presents RUPER-LB, a load balancing system for applications with mixed MPI/multithreading parallelism support with loosely coupling. RUPER-LB focuses on iterative processes running on platforms with variable computational capabilities, such as cloud computing environments. We have shown the capabilities of RUPER-LB using a real world simulation software with MPI and multithreading capabilities. Due to its asynchronous approach, RUPER-LB introduces a negligible overhead on the processing time, making it suitable for applications with few communications. In addition, as RUPER-LB only require periodic reports of thread speeds, it is easily integrable on most applications. 

Future work involves testing RUPER-LB running different kind of applications on both, public and on-premises cloud providers. Also, improving the finish request step to minimize threads waiting time. Finally, extending RUPER-LB to handle the iteration distribution for applications where the iteration migration requires some state transfer.



\bibliography{refs}
\end{document}